\documentclass[11pt]{article}

\usepackage[preprint]{acl}

\usepackage{times}
\usepackage{latexsym}

\usepackage[T1]{fontenc}
\usepackage[utf8]{inputenc}

\usepackage{microtype}

\usepackage{inconsolata}

\usepackage{graphicx}

\usepackage{amsmath}
\usepackage{amssymb}
\usepackage{multirow}
\usepackage{booktabs}

\title{GAMW: Generation-Aware Mel-Domain Watermarking for TTS Traceability}

\author{Yihang Zhang\textsuperscript{1},
  Weicai Yan\textsuperscript{1},
  Qifei Zhang\textsuperscript{1},
  Yangyang Wu\textsuperscript{1},
  Tao Jin\textsuperscript{1}\thanks{Corresponding author.} \\
  \textsuperscript{1}Zhejiang University \\
  \texttt{zhangyihang112@zju.edu.cn},\texttt{jint\_zju@zju.edu.cn}
}

\begin{document}
\maketitle

\begin{abstract}
As text-to-speech (TTS) systems continue to advance, provenance tracking and security verification of synthesized speech have become increasingly important. Audio watermarking provides active provenance by embedding imperceptible, machine-recoverable information into speech. However, existing post-hoc methods typically modify the waveform after synthesis, leaving the watermark vulnerable to compound perturbations. To address this limitation, we propose GAMW, a watermark injection framework integrated into the TTS generation process. Before waveform synthesis, GAMW embeds a content-adaptive multi-bit watermark residual into the generated representation, and its decoding objective combines focal modulation with dynamic per-bit weighting. We train and evaluate GAMW separately on two TTS backbones. With a 16-bit payload, GAMW improves decoding accuracy over the strongest baseline by 9.2 and 23.5 percentage points under individual and compound perturbations, respectively, while maintaining competitive speech quality. Reliable clean-speech decoding is also retained at higher payloads.Code is available at \url{https://github.com/ZConJugaTe/GAMW}.
\end{abstract}
% 随着文本到语音（TTS）系统生成能力不断提升，合成语音的来源追踪与安全验证变得日益重要。音频水印通过在语音中嵌入不可感知且可机器恢复的信息，为合成语音提供主动式内容溯源。然而，现有后处理水印通常在语音生成完成后直接修改波形，在多重扰动下容易破坏水印的鲁棒性。为此，本文提出GAMW，一种面向TTS生成语音过程中的水印注入框架。GAMW在波形合成前向生成过程中嵌入内容自适应的多比特水印残差，其解码目标结合Focal调制与动态逐比特加权。本文在两种TTS模型上分别训练和评估GAMW。实验结果表明，在16比特载荷下，GAMW相较最强基线在单一扰动和复合扰动下分别提高9.2和23.5个百分点，同时保持具有竞争力的语音质量；在更高载荷下，其无扰动解码仍保持可靠。

% 代码可在 \url{https://github.com/ZConJugaTe/GAMW} 获取。

% intro最好写在第二页的第一栏，第二页第二栏写相关工作，占半面。

\section{Introduction}
% 近年来，文本到语音（TTS）技术进展显著，仅需短时参考语音即可合成具有目标音色的高质量语音~\cite{du2024cosyvoice2,cui2025glm}，引发了有关安全和隐私的社会问题，例如深度伪造音频诈骗和版权保护。因此，监管机构必须通过可追溯性方法对合成语音进行监管。被动检测是可追溯性最常用的方法之一，但仅依赖被动检测难以为每段音频提供持久、可验证的来源标识。因此需要一种主动溯源机制。
Recent advances in text-to-speech (TTS) enable high-quality speech synthesis in a target voice from only a short reference utterance~\cite{du2024cosyvoice2,cui2025glm}, but also raise security and privacy concerns, including deepfake audio fraud and challenges in copyright protection~\cite{wenger2021hello,ren2024copyright}. Regulators therefore require traceability mechanisms to oversee synthetic speech. Passive detection is one of the most common approaches to traceability~\cite{lu2023detecting,yan2022initial,xie2023domain}, but alone cannot provide every utterance with a persistent, verifiable source identifier. An active provenance mechanism is therefore needed.

\begin{figure}[t]
\centering
\includegraphics[width=\columnwidth]{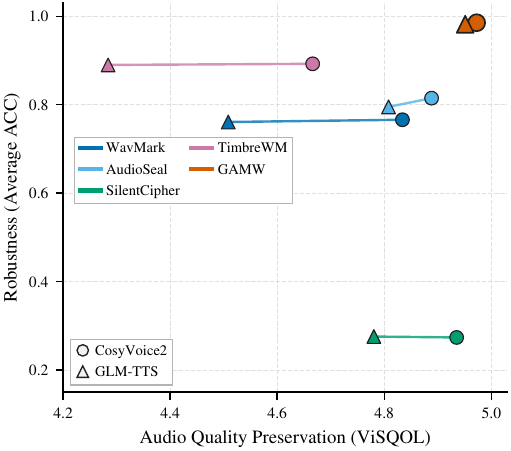}
% 16比特载荷下的音质—鲁棒性权衡。横轴为干净语音与加水印语音之间的ViSQOL，纵轴为单一扰动下的平均解码准确率。圆形和三角形分别表示在CosyVoice2和GLM-TTS上单独训练并评估的结果。
\caption{Audio quality--robustness trade-off with a 16-bit payload. The horizontal axis reports ViSQOL between clean and watermarked speech, and the vertical axis reports the average decoding accuracy under individual perturbations. Circles and triangles denote separately trained and evaluated settings on CosyVoice2 and GLM-TTS, respectively.}
\label{fig:intro_quality_robustness}
\end{figure}

% 音频水印通过嵌入人耳难以察觉但可由算法恢复的消息，为主动溯源提供了可行路径。传统方法依赖手工设计的信号处理规则，其固定的嵌入与检测过程容易受到移除、失同步等攻击，通常需要针对不同攻击设计对策。深度学习方法则联合优化嵌入与提取，学习与语音内容相适应的水印表示。WavMark和AudioSeal是代表性的后处理方法，在语音生成完成后嵌入水印。此类方法部署灵活，但水印写入独立于目标TTS的声学生成过程；现有评测表明，其在水印移除、伪造、基于变换的攻击和神经音频编解码压缩下仍存在鲁棒性局限~\cite{liu2024audiomarkbench,o2025deep,ozer2025comprehensive}。
Audio watermarking enables active provenance by embedding messages that are imperceptible yet algorithmically recoverable. Traditional methods rely on hand-crafted signal-processing rules~\cite{hua2016twenty}, whose fixed embedding and detection procedures are vulnerable to removal, desynchronization, and related attacks and often require attack-specific countermeasures~\cite{arnold2003attacks}. Deep methods instead jointly learn content-adaptive embedding and extraction. WavMark~\cite{chen2023wavmark} and AudioSeal~\cite{roman2024proactive} are representative post-hoc methods that embed watermarks after speech generation. Although flexible to deploy, such methods remain independent of the target TTS acoustic generation process; evaluations reveal robustness limitations under watermark removal, forgery, transformation-based attacks, and neural audio codec compression~\cite{liu2024audiomarkbench,o2025deep,ozer2025comprehensive}.

% 一些研究因而将水印嵌入前移至音频生成过程。GROOT将水印编码到冻结扩散音频生成器的输入潜变量中，使其随生成过程写入输出音频，但这类方案通常依赖特定生成器的潜变量或内部机制。相比之下，mel频谱是许多神经TTS系统中连接声学模型与声码器的中间表示。它既提供显式的时频结构，又采用近似人耳频率感知的非线性mel刻度~\cite{stevens1937scale,davis1980comparison}，使水印能够根据语音内容在时频域分配，并与声学表示共同经过声码器合成。MelShield验证了声码器前mel频谱作为水印载体的可行性，但仍依赖干净mel参考，且未解决音质与鲁棒性之间的权衡。因此，仅选择mel域并不能保证二者，仍需学习与内容相适应的嵌入。
Recent studies have therefore moved watermark insertion into audio generation. GROOT encodes watermarks in the input latent of a frozen diffusion audio generator so that they are written into the output during generation~\cite{liu2024groot}, but such approaches depend on generator-specific latents or mechanisms. In contrast, the mel-spectrogram is a common intermediate representation between the acoustic model and vocoder in neural TTS~\cite{shen2018natural,ren2020fastspeech}. It provides an explicit time--frequency structure and uses a nonlinear mel scale that approximates human frequency perception~\cite{stevens1937scale,davis1980comparison}, allowing watermarks to follow speech content across time and frequency and be rendered with the acoustic representation by the vocoder. MelShield demonstrates the feasibility of pre-vocoder mel watermarking, but requires a clean mel reference and leaves the fidelity--robustness trade-off unresolved~\cite{jin2026melshield}. Thus, choosing the mel domain alone guarantees neither property; content-adaptive embedding remains necessary.

% 为此，本文提出Generation-Aware Mel-Domain Watermarking（GAMW），一种面向TTS生成语音溯源的学习式mel频谱水印框架。在TTS生成的mel频谱上嵌入内容自适应的多比特水印残差，并通过冻结声码器将其渲染为波形；提取器从待验证语音恢复水印载荷，无需原始mel参考。训练目标由解码损失与mel保真度损失组成：解码损失结合Focal调制与动态逐比特加权，分别关注低置信度的样本—比特实例和当前批次中恢复较弱的比特位置；mel保真度损失约束加水印频谱与干净频谱之间的偏差，从而协调水印恢复与频谱保真度。训练中TTS主干保持冻结。如图~\ref{fig:intro_quality_robustness}所示，GAMW在音质与扰动后水印恢复之间取得了良好的经验性权衡。
To address this challenge, we propose Generation-Aware Mel-Domain Watermarking (GAMW), a learned mel-spectrogram watermarking framework for TTS provenance. GAMW embeds a content-adaptive multi-bit residual into the TTS-generated mel-spectrogram, which the frozen vocoder renders into a waveform. The extractor recovers the payload from speech under verification without the original mel reference. Its objective comprises decoding and mel-fidelity losses. The former combines focal modulation~\cite{lin2017focal} with dynamic per-bit weighting to emphasize low-confidence sample--bit instances and weak bit positions in the current mini-batch, respectively; the latter constrains deviations between watermarked and clean spectrograms, balancing recovery and spectral fidelity. The TTS backbone remains frozen during training. As shown in Fig.~\ref{fig:intro_quality_robustness}, GAMW achieves a favorable empirical trade-off between speech quality and post-perturbation watermark recovery.

% 本文的主要贡献：
Our main contributions are as follows:

\begin{itemize}
    % 提出一种面向TTS生成语音溯源的学习式mel频谱水印框架，在冻结TTS主干的条件下，将多比特水印嵌入mel频谱，使合成音频携带水印。
    \item We propose a learned mel-spectrogram watermarking framework for TTS provenance that embeds multi-bit watermarks with the TTS backbone frozen, allowing synthesized speech to carry the watermark.

    % 针对多比特水印任务构建一种结合Focal调制与动态逐比特加权的解码目标，以缓解样本—比特实例与比特位置之间的恢复难度不均衡。
    \item We design a decoding objective combining focal modulation and dynamic per-bit weighting to address unequal recovery difficulty across sample--bit instances and bit positions.

    % 在两个TTS模型上进行实验验证，实验覆盖音质、载荷容量、单一与复合扰动，以及嵌入域和损失函数消融。
    \item We validate GAMW on two TTS backbones across audio quality, payload capacity, individual and compound perturbations, and embedding-domain and loss-function ablations.
\end{itemize}

\section{Related Work}
\subsection{Text-to-Speech Models}
% 神经TTS通常将声学表示生成与波形合成分开。Tacotron 2以自回归方式预测mel频谱，FastSpeech及FastSpeech 2采用并行非自回归生成，WaveNet、HiFi-GAN和HiFTNet等神经声码器则将声学表示转换为波形。随着离散语音表示的发展，VALL-E系列将零样本TTS建模为codec token上的条件语言建模。CosyVoice系列与GLM-TTS均包含语音token生成、基于流的mel频谱生成和神经声码器，并保留mel到波形的生成阶段。GAMW在mel域上嵌入水印，使得输出音频自然携带水印。
Neural TTS commonly separates acoustic-representation generation from waveform synthesis. Tacotron~2 predicts mel-spectrograms autoregressively, FastSpeech and FastSpeech~2 use parallel non-autoregressive generation~\cite{shen2018natural,ren2019fastspeech,ren2020fastspeech}, and vocoders such as WaveNet, HiFi-GAN, and HiFTNet convert acoustic representations into waveforms~\cite{van2016wavenet,kong2020hifi,li2023hiftnet}. With discrete speech representations, the VALL-E series formulates zero-shot TTS as conditional language modeling over codec tokens~\cite{wang2023neural,chen2024vall}. Both the CosyVoice series and GLM-TTS employ speech-token generation, flow-based mel-spectrogram generation, and neural vocoding, retaining a mel-to-waveform synthesis stage~\cite{du2024cosyvoice,du2024cosyvoice2,du2025cosyvoice,cui2025glm}. GAMW embeds watermarks in the mel domain, allowing the synthesized speech to carry the watermark after waveform generation.

\subsection{Audio Watermarking}
% 传统音频水印主要依赖人工设计的信号处理方法，包括最低有效位修改、回声隐藏、扩频和量化索引调制，并通过具体的嵌入规则在音质、容量和鲁棒性之间进行权衡。
Traditional audio watermarking relies mainly on hand-crafted signal-processing techniques, including least-significant-bit modification~\cite{cvejic2004increasing}, echo hiding~\cite{cho2003echo}, spread spectrum~\cite{kirovski2003spread}, and quantization index modulation~\cite{chen2001quantization}. Their embedding rules are designed to trade off fidelity, capacity, and robustness.

% 深度学习水印通过联合优化嵌入与提取来学习与内容适配的水印模式。WavMark使用可逆神经网络完成水印嵌入与提取，AudioSeal支持局部水印检测，SilentCipher将心理声学约束与伪可微压缩相结合，TimbreWM则提出音色水印技术以抵御语音克隆攻击。相关研究还考虑了重录音失真和跨注意力恢复。上述方法易于应用到已有音频，但水印写入独立于目标TTS的生成过程。另一些方法将水印前移至生成阶段：GROOT在扩散模型的潜变量中编码水印，TraceableSpeech在TTS生成时逐帧写入水印，SOLIDO通过低秩适配修改扩散模型。MelShield进一步利用声码器前的mel表示，但验证依赖干净mel参考及元数据。GAMW同样在mel接口嵌入水印，同时在冻结TTS主干的条件下实现无需原始mel参考的多比特恢复。
Deep watermarking instead learns content-adaptive embedding and recovery jointly. WavMark uses invertible networks for watermark embedding and extraction~\cite{chen2023wavmark}, while AudioSeal supports localized watermark detection~\cite{roman2024proactive}. SilentCipher combines psychoacoustic constraints with pseudo-differentiable compression~\cite{singh2024silentcipher}, and TimbreWM introduces timbre watermarking to resist voice-cloning attacks~\cite{liu2023detecting}. Related studies also consider re-recording distortions and cross-attention-based recovery~\cite{pavlovic2022robust,liu2023dear,liu2025xattnmark}. These methods can be applied to existing audio, but watermark insertion remains separate from the target TTS generation process. Other methods move insertion into generation: GROOT encodes watermarks in diffusion latents~\cite{liu2024groot}, TraceableSpeech writes them frame by frame during TTS generation~\cite{zhou2024traceablespeech}, and SOLIDO modifies diffusion models through low-rank adaptation~\cite{li2025solido}. MelShield further uses a pre-vocoder mel representation, but requires a clean mel reference and metadata for verification~\cite{jin2026melshield}. GAMW also embeds watermarks at the mel interface while enabling reference-free multi-bit recovery with a frozen TTS backbone.

\section{Method}
\subsection{Overall Framework}
\begin{figure*}[t]
\centering
\includegraphics[width=\textwidth]{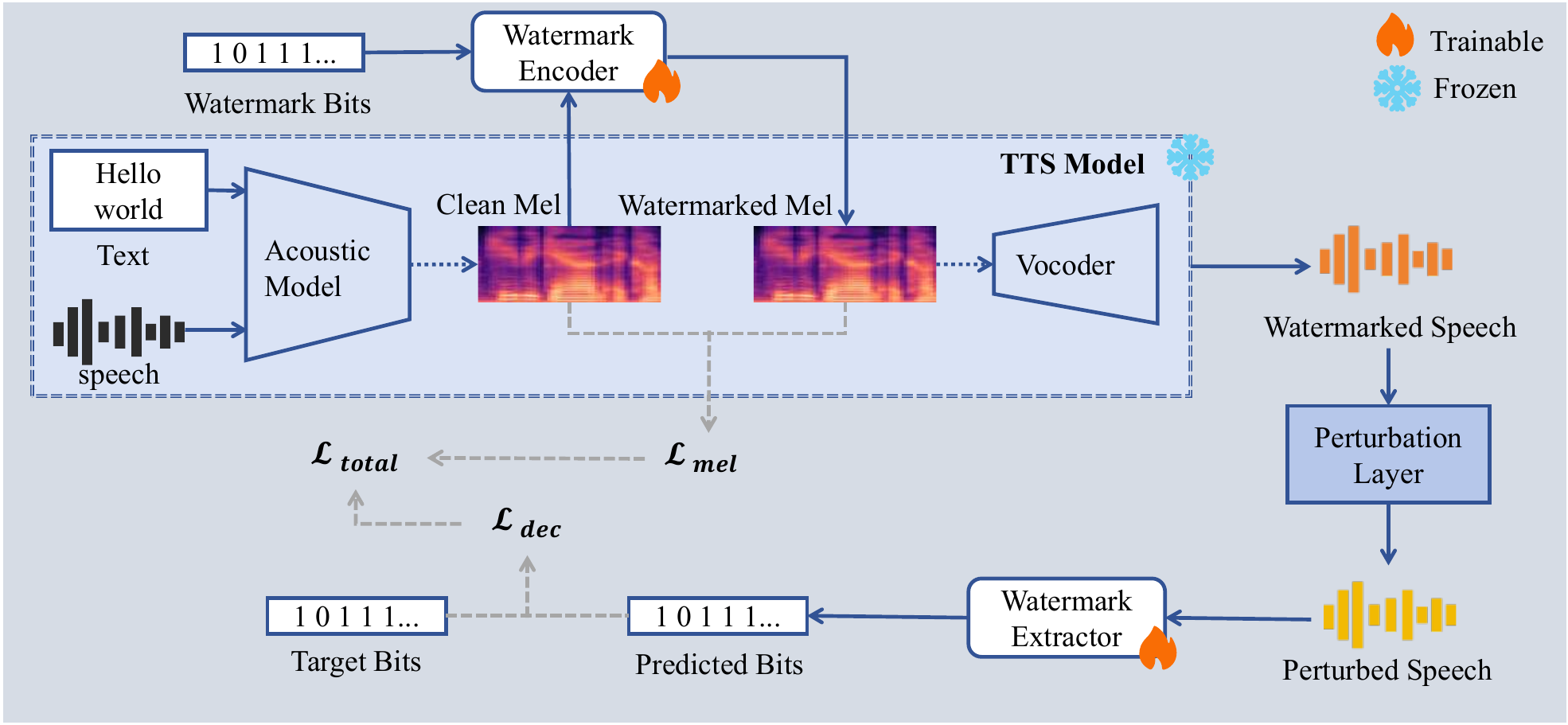}
% GAMW总体框架。可训练的水印编码器将多比特载荷嵌入冻结TTS模型生成的干净mel频谱，冻结声码器将加水印mel频谱合成为语音。训练阶段对语音施加扰动，并由可训练的水印提取器恢复载荷；总目标联合优化解码准确性与mel域保真度。
\caption{Overview of GAMW. The trainable watermark encoder embeds a multi-bit payload into the clean mel-spectrogram generated by the frozen TTS model, and the frozen vocoder synthesizes the watermarked speech. During training, perturbations are applied before payload recovery by the trainable extractor, while the objective jointly optimizes decoding accuracy and mel-domain fidelity.}
\label{fig:watermark}
\end{figure*}

% GAMW的总体框架如图~\ref{fig:watermark}所示。给定输入文本，预训练TTS模型生成干净mel频谱$M \in \mathbb{R}^{C \times T}$，其中$C$和$T$分别表示mel频带数和时间帧数。水印编码器将$K$比特二进制序列$w \in \{0,1\}^K$嵌入$M$，得到加水印mel频谱$\tilde{M}$，经冻结的声码器得到含水印语音。训练阶段对含水印语音随机施加扰动，并由水印提取器从扰动后的波形中恢复载荷。整个过程中，TTS主干与声码器参数均保持冻结，仅联合优化水印编码器和提取器。
As illustrated in Fig.~\ref{fig:watermark}, given an input text, a pre-trained TTS model generates a clean mel-spectrogram $M \in \mathbb{R}^{C \times T}$, where $C$ and $T$ denote the numbers of mel bands and time frames, respectively. The watermark encoder embeds a $K$-bit binary sequence $w \in \{0,1\}^K$ into $M$ to obtain the watermarked mel-spectrogram $\tilde{M}$, which is synthesized into watermarked speech by the frozen vocoder. During training, random perturbations are applied to this speech, and the watermark extractor recovers the payload from the perturbed waveform. The TTS backbone and vocoder remain frozen throughout, while only the watermark encoder and extractor are jointly optimized.

% 与后处理方式不同，GAMW在波形生成前作用于的mel声学表示，使水印与语音内容共同经过声码器渲染并进入最终波形。mel频谱的显式时频结构使编码器能够依据局部声学内容分配水印信号，从而在控制感知失真的同时维持可恢复的嵌入强度。这一设计为协调语音质量与对扰动的鲁棒性提供了可学习的路径，其实际效果将在实验中验证。
Unlike post-processing methods, GAMW operates on the mel acoustic representation before waveform generation, allowing the watermark and speech content to be jointly rendered into the final waveform by the vocoder. The explicit time--frequency structure of the mel-spectrogram enables content-aware allocation of watermark signals, preserving recoverable embedding strength while limiting perceptual distortion. This design provides a learnable way to balance speech quality and perturbation robustness, as evaluated in our experiments.

\subsection{Watermark Encoder}
\begin{figure}[htbp]
\centering
\includegraphics[width=\columnwidth]{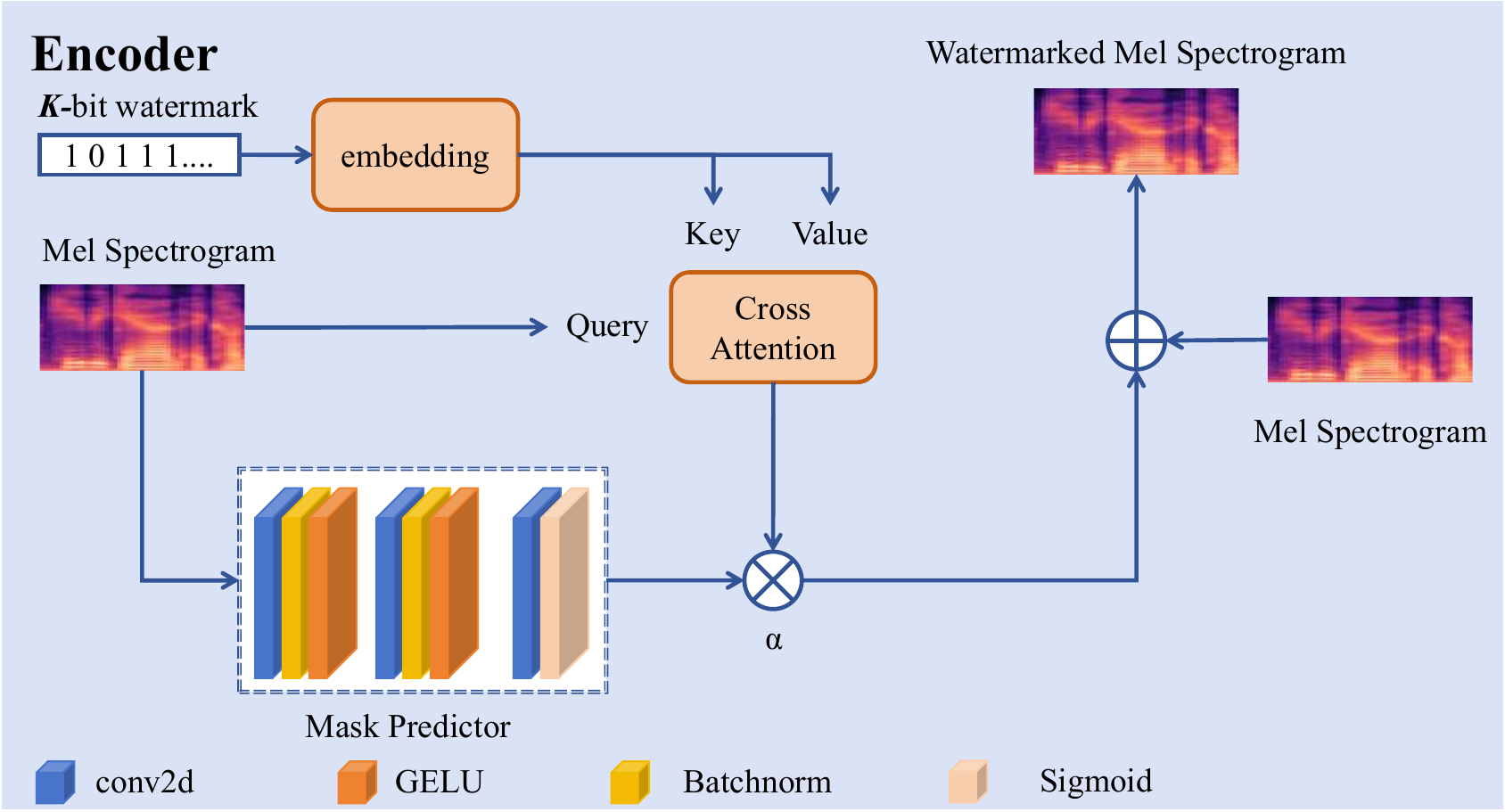}
% 水印编码器。交叉注意力根据mel内容融合水印表示，掩码预测器与可学习全局强度共同调节水印残差，随后将其加入干净mel频谱。
\caption{Watermark encoder. Cross-attention fuses the watermark representation with the mel content, while the mask predictor and learnable global strength jointly modulate the watermark residual before it is added to the clean mel-spectrogram.}
\label{fig:encoder}
\end{figure}

% 水印编码器如图~\ref{fig:encoder}所示，编码器将水印二进制序列$w$写入干净mel频谱$M$并输出带水印mel频谱$\tilde{M}$，使后续合成的语音携带相应水印。
As illustrated in Fig.~\ref{fig:encoder}, the watermark encoder embeds the binary sequence $w$ into the clean mel-spectrogram $M$ and outputs the watermarked mel-spectrogram $\tilde{M}$, enabling the subsequently synthesized speech to carry the corresponding watermark.

% 水印比特经可学习的嵌入层映射后与位置编码相加，得到同时表征比特取值与位置顺序的水印表示$E_w \in \mathbb{R}^{K \times d_{\text{emb}}}$，其中$d_{\text{emb}}$为嵌入维度，并设置为与mel频带数$C$相同。
The watermark bits are mapped through a learnable embedding layer and combined with positional encodings, yielding $E_w \in \mathbb{R}^{K \times d_{\text{emb}}}$, which represents both bit values and their positions. We set the embedding dimension $d_{\text{emb}}$ equal to the number of mel bands $C$.

% 多头交叉注意力随后将水印表示映射至mel频谱空间。mel帧作为查询，$E_w$作为键和值，使各时间帧结合自身声学内容聚合水印信息，生成内容相关的原始水印残差$S_{\text{raw}} \in \mathbb{R}^{C \times T}$：
Multi-head cross-attention~\cite{vaswani2017attention} then maps the watermark representation into the mel-spectrogram space. Mel frames serve as queries, while $E_w$ provides the keys and values, allowing each frame to aggregate watermark information conditioned on its acoustic content and producing a content-dependent raw residual $S_{\text{raw}} \in \mathbb{R}^{C \times T}$:
\begin{equation}
S_{\text{raw}} = \text{CrossAttn}\bigl(Q=M^{\top},\, K=E_w,\, V=E_w\bigr)^{\top}
\end{equation}

% 为使水印分布适应语音内容，我们使用轻量级的二维卷积网络根据$M$预测时频软掩码$M_{\text{mask}}$，并依据局部声学结构调节各时频单元的相对水印强度；可学习的全局因子$\alpha$则控制整体嵌入强度。二者共同调节水印的空间分布与幅度，使编码器在可恢复性与频谱保真度之间进行自适应权衡。加水印mel频谱表示为：
To adapt the watermark distribution to speech content, a lightweight 2D convolutional network predicts a time--frequency soft mask $M_{\text{mask}}$ from $M$ and adjusts the relative watermark strength of each time--frequency bin according to local acoustic structure. A learnable global factor $\alpha$ controls the overall embedding strength. Together, they regulate the spatial distribution and magnitude of the watermark, enabling the encoder to adaptively balance recoverability and spectral fidelity. The watermarked mel-spectrogram is given by
\begin{equation}
\tilde{M} = M + \alpha \cdot M_{\text{mask}} \odot S_{\text{raw}}
\end{equation}
% 其中，$\odot$表示逐元素乘法。该残差形式保留原始mel频谱的直接路径，并将水印表示为受时频掩码和全局强度共同调节的加性修改。
where $\odot$ denotes element-wise multiplication. This residual formulation preserves a direct path for the original mel-spectrogram and expresses the watermark as an additive modification jointly regulated by the mask and global strength.

\subsection{Watermark Extractor}
\begin{figure}[htbp]
\centering
\includegraphics[width=\columnwidth]{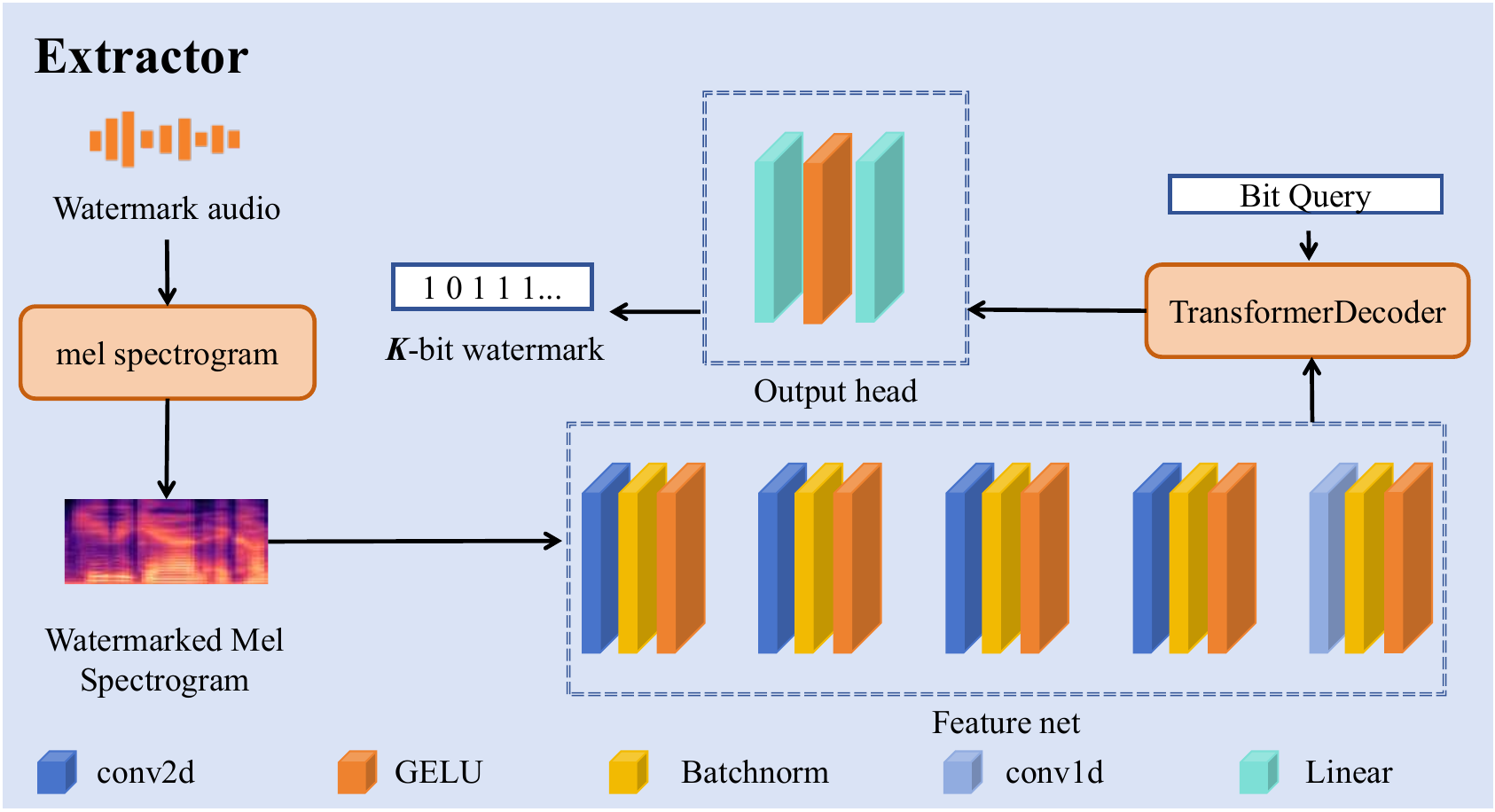}
% 水印提取器。输入语音首先转换为mel频谱，随后由卷积特征网络提取时频特征，并通过带可学习查询的Transformer解码器和输出头恢复多比特载荷。
\caption{Watermark extractor. The input speech is converted into a mel-spectrogram, from which a convolutional feature network extracts time--frequency features. A Transformer decoder with a learnable query aggregates the features, and an output head recovers the multi-bit payload.}
\label{fig:extractor}
\end{figure}
% 水印提取器从可能经过扰动的音频中恢复载荷。输入波形通过参数固定的mel变换转换为时频表示，随后由二维卷积骨干沿频率轴逐步下采样，同时保留时间结构，得到特征序列$F \in \mathbb{R}^{T' \times d_{\text{feat}}}$。Transformer解码器以所有样本共享的可学习查询$Q \in \mathbb{R}^{1 \times d_{\text{feat}}}$为输入，从$F$中聚合全局水印证据：
The watermark extractor recovers the payload from audio that may have undergone perturbations. A fixed-parameter mel transform converts the input waveform into a time--frequency representation. A 2D convolutional backbone then progressively downsamples the frequency axis while preserving temporal structure, producing a feature sequence $F \in \mathbb{R}^{T' \times d_{\text{feat}}}$. The Transformer decoder uses a learnable query $Q \in \mathbb{R}^{1 \times d_{\text{feat}}}$ shared across all samples to aggregate global watermark evidence from $F$:
\begin{equation}
h = \text{TransformerDecoder}\bigl(\text{tgt}=Q,\; \text{memory}=F\bigr)
\end{equation}
% 聚合表示$h \in \mathbb{R}^{1 \times d_{\text{feat}}}$经预测头映射为$K$个水印比特的概率$p \in [0,1]^K$，并通过阈值化得到最终的二进制载荷。该过程仅使用待验证音频，无需原始语音或干净mel频谱作为参考。
The prediction head maps the aggregated representation $h \in \mathbb{R}^{1 \times d_{\text{feat}}}$ to the probabilities $p \in [0,1]^K$ of the $K$ watermark bits, which are thresholded to obtain the final binary payload. This process uses only the suspect audio and requires neither the original speech nor the clean mel-spectrogram as a reference.

\subsection{Training Objective}
% 训练目标由解码损失与mel域保真度损失加权构成：
The training objective is a weighted combination of the decoding loss and the mel-domain fidelity loss:
\begin{equation}
\mathcal{L}_{\text{total}} = \lambda_{\text{dec}} \cdot \mathcal{L}_{\text{dec}} + \lambda_{\text{mel}} \cdot \mathcal{L}_{\text{mel}}
\end{equation}
% 其中，$\lambda_{\text{dec}}$和$\lambda_{\text{mel}}$为平衡载荷恢复与频谱保真度的超参数。
where $\lambda_{\text{dec}}$ and $\lambda_{\text{mel}}$ are hyperparameters that balance payload recovery and spectral fidelity.

% 多比特水印的恢复难度体现在两个层面：同一批次中的样本—比特实例具有不同的预测置信度，不同比特位置也可能呈现不同的恢复难度。为同时处理这两类不均衡，本文在Focal二元交叉熵的基础上引入动态逐比特权重。Focal调制降低高置信度易分类实例的损失贡献，逐比特权重则根据当前批次中各位置的解码准确率，提高恢复较困难位置的优化权重：
Multi-bit watermark recovery is imbalanced at two levels: sample--bit instances within a mini-batch have different prediction confidence, and bit positions may exhibit different recovery difficulty. To address both forms of imbalance, we augment focal binary cross-entropy~\cite{lin2017focal} with dynamic per-bit weighting. Focal modulation reduces the loss contribution of high-confidence, easy instances, while the per-bit weights emphasize harder positions according to their decoding accuracy in the current mini-batch:
\begin{equation}
\mathcal{L}_{\text{dec}} = \frac{1}{BK} \sum_{i=1}^{B}\sum_{k=1}^{K} \omega_k (1 - p_{t,ik})^{\gamma} \text{BCE}(p_{ik}, w_{ik})
\end{equation}
\begin{equation}
\omega_k = 1 + \lambda_w \cdot (1 - \text{acc}_k)
\end{equation}
% 其中，$B$为批次大小，$p_{ik}$和$w_{ik}$分别表示第$i$个样本中第$k$个比特的预测概率与真实标签，$p_{t,ik}=p_{ik}w_{ik}+(1-p_{ik})(1-w_{ik})$为模型赋予正确类别的概率。$\text{acc}_k$由阈值化预测在当前批次内计算，并在构造权重时停止梯度传播。超参数$\gamma$控制对易分类实例的抑制程度，$\lambda_w$控制逐比特重加权强度。因此，两种权重分别作用于实例级置信度和比特位置级难度。通过将优化重点同时转向低置信度实例和当前批次内恢复较弱的比特位置，该设计有助于加快解码任务的学习。
Here, $B$ denotes the mini-batch size; $p_{ik}$ and $w_{ik}$ are the predicted probability and ground-truth label for bit $k$ of sample $i$, respectively; and $p_{t,ik}=p_{ik}w_{ik}+(1-p_{ik})(1-w_{ik})$ is the probability assigned to the correct class. The accuracy $\text{acc}_k$ is computed from thresholded predictions over the current mini-batch and detached when constructing the weights. The hyperparameter $\gamma$ controls the suppression of easy instances, while $\lambda_w$ controls the strength of per-bit reweighting. The two weighting terms therefore act on instance-level confidence and bit-position-level difficulty, respectively. By directing optimization toward both low-confidence instances and weakly recovered bit positions within the current mini-batch, this design facilitates faster learning of the decoding task.

% mel域保真度损失采用均方误差，直接约束加水印mel频谱$\tilde{M}$与原始干净mel频谱$M$之间的偏差：
The mel-domain fidelity loss uses mean squared error to directly constrain the deviation between the watermarked mel-spectrogram $\tilde{M}$ and the original clean mel-spectrogram $M$:
\begin{equation}
\mathcal{L}_{\text{mel}} = \frac{1}{C T} \bigl\|\tilde{M} - M\bigr\|_{\mathrm{F}}^2
\end{equation}

\section{Experiments}
% 本节从语音质量、载荷容量、鲁棒性和消融实验四个方面评估GAMW。
We evaluate GAMW in terms of audio quality, payload capacity, robustness, and ablation studies.

\subsection{Experimental Setup}
% 数据集。实验数据由LibriSpeech和LJSpeech两部分构成。测试数据分别来自LibriSpeech的官方测试集和从LJSpeech中人工划分的测试集，所有测试样本在训练阶段均不可见。
\textbf{Datasets.} The experimental data consist of LibriSpeech~\cite{panayotov2015librispeech} and LJSpeech~\cite{ljspeech17}. We use the official LibriSpeech test set and a manually held-out test split from LJSpeech. All test samples are excluded from training.

% 评估指标。语音质量采用PESQ、ViSQOL和MOS评估。PESQ和ViSQOL为全参考指标，通过比较加水印语音与对应的干净语音衡量水印引入的感知退化。MOS由UTMOS模型预测，用于估计加水印语音的整体感知质量。水印恢复采用逐比特解码准确率（ACC）评估。所有指标均为数值越高越好。
\textbf{Evaluation Metrics.} Speech quality is evaluated using PESQ~\cite{rix2001perceptual}, ViSQOL~\cite{hines2015visqol}, and MOS. PESQ and ViSQOL are full-reference metrics that compare watermarked speech with its clean synthetic counterpart to quantify watermark-induced perceptual degradation. MOS is estimated by the no-reference UTMOS model~\cite{saeki2022utmos} and reflects overall perceptual quality. Watermark recovery is measured by bitwise decoding accuracy (ACC). Higher values are better for all metrics.

% 基座模型。我们分别在CosyVoice2和GLM-TTS上训练GAMW，并采用相同的数据与训练协议。具体实现设置见附录~\ref{app:implementation_details}。
\textbf{TTS Backbones.} We train GAMW separately on CosyVoice2~\cite{du2024cosyvoice2} and GLM-TTS~\cite{cui2025glm} using the same data and training protocol. Detailed implementation settings are provided in Appendix~\ref{app:implementation_details}.

% 基线方法。我们评测四种语音水印基线，包括WavMark、AudioSeal、SilentCipher和TimbreWM，其水印容量分别为32、16、40和10比特。
\textbf{Baselines.} We evaluate four speech watermarking baselines: WavMark~\cite{chen2023wavmark}, AudioSeal~\cite{roman2024proactive}, SilentCipher~\cite{singh2024silentcipher}, and TimbreWM~\cite{liu2023detecting}, with payloads of 32, 16, 40, and 10 bits, respectively.

\subsection{Audio Quality Evaluation}
% 本实验在无扰动条件下评估水印嵌入对合成语音质量的影响，并报告无扰动水印解码准确率作为参考。
We evaluate the impact of watermark embedding on synthetic speech quality under clean conditions and report clean watermark decoding accuracy for reference.

\begin{table}[htbp]
\centering
\small
\setlength{\tabcolsep}{4.5pt}
\caption{Audio quality comparison under the no-perturbation condition. MOS is predicted by UTMOS. Higher is better for all metrics.}
\label{tab:quality}
\begin{tabular}{lcccc}
\hline
\textbf{Method} & \textbf{PESQ} & \textbf{ViSQOL} & \textbf{MOS} & \textbf{ACC} \\
\hline
\multicolumn{5}{c}{\textbf{CosyVoice2}} \\
\hline
WavMark              & 4.3008 & 4.8340 & 4.0628 & 1.0000 \\
AudioSeal            & 4.3672 & 4.8884 & 4.0592 & 1.0000 \\
SilentCipher         & 4.5701 & 4.9352 & 4.0745 & 0.8654 \\
TimbreWM             & 3.8464 & 4.6662 & 3.9354 & 1.0000 \\
GAMW (ours)          & 4.3001 & 4.9726 & 4.0775 & 1.0000 \\
\hline
\multicolumn{5}{c}{\textbf{GLM-TTS}} \\
\hline
WavMark              & 4.2259 & 4.5085 & 3.7017 & 1.0000 \\
AudioSeal            & 4.3529 & 4.8080 & 3.6893 & 1.0000 \\
SilentCipher         & 4.5729 & 4.7806 & 3.7197 & 0.8512 \\
TimbreWM             & 3.8114 & 4.2841 & 3.5633 & 1.0000 \\
GAMW (ours)          & 4.2970 & 4.9512 & 3.7245 & 1.0000 \\
\hline
\end{tabular}
\end{table}
% 表~\ref{tab:quality}显示，GAMW在两个TTS骨干上均取得最高的ViSQOL，同时保持具有竞争力的PESQ。其MOS也均高于四种基线，表明水印引入的感知失真有限，并能较好地保持整体语音质量。无扰动下，GAMW在两个骨干上的ACC均为1.0000；除SilentCipher外，其余基线也达到1.0000。
Table~\ref{tab:quality} shows that GAMW achieves the highest ViSQOL on both TTS backbones while maintaining competitive PESQ. Its MOS also exceeds those of all four baselines on both backbones, indicating limited watermark-induced perceptual degradation and well-preserved overall speech quality. GAMW achieves an ACC of 1.0000 on both backbones under clean conditions; all baselines except SilentCipher also achieve 1.0000.

\subsection{Capacity Evaluation}
% 本实验在无扰动条件下测试不同水印容量的影响，包含语音质量与解码准确率。
We evaluate the effects of different watermark payload sizes under clean conditions in terms of speech quality and decoding accuracy.

\begin{table}[htbp]
\centering
\small
\caption{Capacity evaluation at different watermark lengths. MOS is predicted by UTMOS. Higher is better for all metrics.}
\label{tab:capacity}
\begin{tabular}{ccccc}
\hline
\textbf{$K$ (bits)} & \textbf{PESQ} & \textbf{ViSQOL} & \textbf{MOS} & \textbf{ACC} \\
\hline
\multicolumn{5}{c}{\textbf{CosyVoice2}} \\
\hline
16  & 4.3001 & 4.9726 & 4.0775 & 1.0000 \\
32  & 4.2518 & 4.9614 & 4.0766 & 0.9976 \\
50  & 4.2332 & 4.9562 & 4.0763 & 0.9992 \\
100 & 4.0607 & 4.9205 & 4.0266 & 0.9978 \\
\hline
\multicolumn{5}{c}{\textbf{GLM-TTS}} \\
\hline
16  & 4.2970 & 4.9512 & 3.7245 & 1.0000 \\
32  & 4.2042 & 4.9316 & 3.7223 & 0.9996 \\
50  & 4.1435 & 4.9122 & 3.7143 & 0.9991 \\
100 & 4.0676 & 4.8823 & 3.6708 & 0.9903 \\
\hline
\end{tabular}
\end{table}

% 如表~\ref{tab:capacity}所示，随着载荷从16比特增加至100比特，两种TTS骨干上的PESQ、ViSQOL和MOS整体呈缓慢下降趋势，表明更高容量会增加水印引入的失真。尽管如此，在100比特下，CosyVoice2和GLM-TTS的ACC仍分别达到0.9978和0.9903，同时PESQ均高于4.06、ViSQOL均高于4.88。结果表明，GAMW在无扰动条件下能够将载荷扩展至100比特，并保持较高的解码可靠性和有限的感知质量损失。
As shown in Table~\ref{tab:capacity}, PESQ, ViSQOL, and MOS generally decrease gradually as the payload increases from 16 to 100 bits on both TTS backbones, indicating greater watermark-induced distortion at higher capacities. Nevertheless, with a 100-bit payload, GAMW retains ACCs of 0.9978 and 0.9903 on CosyVoice2 and GLM-TTS, respectively, while PESQ remains above 4.06 and ViSQOL above 4.88. These results show that GAMW can scale to a 100-bit payload under clean conditions while maintaining reliable decoding and limited perceptual degradation.

\subsection{Robustness Evaluation}
% 鲁棒性实验采用16比特主设置，分别考察单一扰动和复合扰动。单一扰动参考AudioMarkBench的评测类别进行选择；复合扰动按预设顺序串联多种处理，以模拟复杂的传播与编辑过程。具体扰动设置见附录~\ref{app:perturbation_settings}。
Robustness is evaluated using the primary 16-bit setting under individual and compound perturbations. Individual perturbations are selected based on the categories in AudioMarkBench~\cite{liu2024audiomarkbench}, while compound perturbations chain multiple operations in predefined orders to simulate complex transmission and editing. Detailed perturbation settings are provided in Appendix~\ref{app:perturbation_settings}.

\subsubsection{Single Perturbation}
\begin{table*}[t]
\centering
\small
\setlength{\tabcolsep}{3pt}
\caption{Decoding accuracy under individual perturbations on CosyVoice2 with a 16-bit payload. The Average row is computed over all 15 conditions. Complete results for both backbones are provided in Appendix Table~\ref{tab:single_attack_full}. Higher is better.}
\label{tab:single_attack}
\begin{tabular}{lccccc}
\hline
\textbf{Perturbation} & \textbf{WavMark} & \textbf{AudioSeal} & \textbf{SilentCipher} & \textbf{TimbreWM} & \textbf{GAMW} \\
\hline
\multicolumn{6}{c}{\textbf{CosyVoice2}} \\
\hline
Gaussian noise (SNR = 10 dB)       & 0.5008 & 0.6711 & 0.1669 & 0.6777 & 0.9837 \\
Gaussian noise (SNR = 20 dB)       & 0.5946 & 0.8911 & 0.1751 & 0.9054 & 0.9977 \\
Gaussian noise (SNR = 30 dB)       & 0.8572 & 0.9876 & 0.2370 & 0.9922 & 0.9996 \\
Background noise (SNR = 10 dB)     & 0.6176 & 0.7256 & 0.2343 & 0.7224 & 0.9663 \\
Background noise (SNR = 20 dB)     & 0.8580 & 0.8991 & 0.2635 & 0.8918 & 0.9963 \\
Background noise (SNR = 30 dB)     & 0.9817 & 0.9842 & 0.3327 & 0.9900 & 0.9995 \\
Low-pass filtering                 & 0.9875 & 1.0000 & 0.8191 & 0.9976 & 0.9997 \\
High-pass filtering                & 0.9999 & 0.3253 & 0.3108 & 0.9672 & 0.9575 \\
MP3 compression                    & 0.7123 & 0.9850 & 0.2273 & 0.9858 & 0.9716 \\
Opus compression                   & 0.7809 & 0.8323 & 0.2115 & 0.9822 & 0.9709 \\
EnCodec~\cite{defossez2022high}    & 0.4454 & 0.3691 & 0.1698 & 0.6045 & 0.9670 \\
Smoothing                          & 0.8953 & 0.8750 & 0.2701 & 0.9081 & 0.9951 \\
Echo                               & 0.8332 & 0.9608 & 0.2012 & 0.9988 & 0.9967 \\
Quantization                       & 0.4673 & 0.7189 & 0.1835 & 0.7637 & 0.9816 \\
Time stretching                    & 0.9600 & 0.9980 & 0.3081 & 0.9998 & 0.9981 \\
\hline
Average                            & 0.7661 & 0.8149 & 0.2740 & 0.8925 & 0.9854 \\
\hline
\end{tabular}
\end{table*}
% 表~\ref{tab:single_attack}展示GAMW和四种基线方法在CosyVoice2上全部15种单一扰动下的鲁棒性，平均值由15种扰动条件等权计算。CosyVoice2和GLM-TTS的完整结果见附录表~\ref{tab:single_attack_full}。
Table~\ref{tab:single_attack} reports all 15 individual-perturbation results for GAMW and four baselines on CosyVoice2, with the average computed equally over the 15 conditions. Complete results for CosyVoice2 and GLM-TTS are provided in Appendix Table~\ref{tab:single_attack_full}.

% 如表~\ref{tab:single_attack}所示，GAMW在CosyVoice2上的平均ACC达到0.9854，并在15种扰动条件中的9种取得最佳结果。附录表~\ref{tab:single_attack_full}显示GLM-TTS呈现一致趋势：GAMW的平均ACC为0.9820，同样在9种条件下排名第一。其优势主要体现在更具挑战性的扰动上：在全部加性噪声条件下，GAMW均排名第一；对于EnCodec重合成，其在CosyVoice2和GLM-TTS上的ACC仍分别达到0.9670和0.9420，而最强基线仅为0.6045和0.5971；在平滑和量化下也保持领先。在其余滤波、传统压缩、回声和时间拉伸条件下，个别基线略占优势，但GAMW仍分别保持至少0.9575和0.9622的ACC。上述结果表明，GAMW在保持常见处理下竞争力的同时，对噪声、量化和EnCodec重合成表现出更强的整体鲁棒性。
As shown in Table~\ref{tab:single_attack}, GAMW achieves an average ACC of 0.9854 on CosyVoice2 and ranks first in nine of the fifteen conditions. Appendix Table~\ref{tab:single_attack_full} shows the same trend on GLM-TTS, where GAMW reaches an average ACC of 0.9820 and again ranks first in nine conditions. Its advantage is concentrated on the more challenging conditions: GAMW ranks first under every additive-noise setting and retains ACCs of 0.9670 and 0.9420 after EnCodec re-synthesis on CosyVoice2 and GLM-TTS, respectively, compared with 0.6045 and 0.5971 for the strongest baseline. It also leads under smoothing and quantization. Under the remaining filtering, conventional compression, echo, and time-stretching conditions, individual baselines perform slightly better, but GAMW still maintains ACCs of at least 0.9575 and 0.9622 on the two backbones. These results show that GAMW combines competitive performance under common processing with stronger overall robustness to noise, quantization, and EnCodec re-synthesis.

\subsubsection{Compound Perturbation}
\begin{figure*}[t]
\centering
\includegraphics[width=\textwidth]{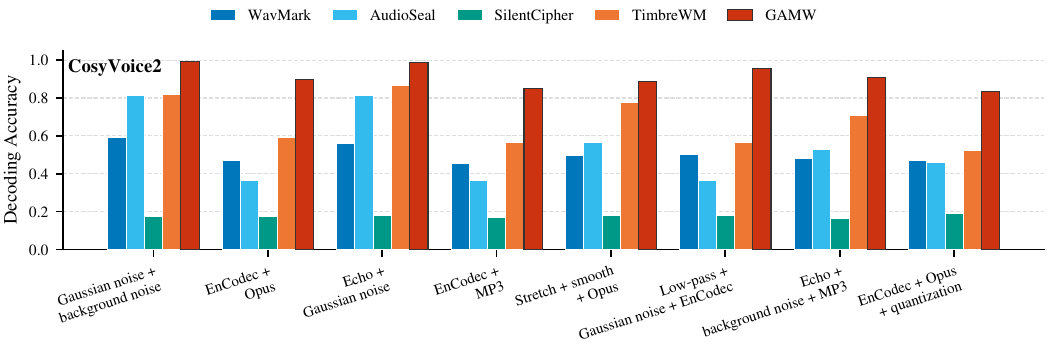}
% 16比特载荷下CosyVoice2上八种复合扰动的水印解码准确率。完整的CosyVoice2和GLM-TTS数值见附录表~\ref{tab:compound_attack}。数值越高越好。
\caption{Watermark decoding accuracy under eight compound perturbations on CosyVoice2 with a 16-bit payload. Complete numerical results for CosyVoice2 and GLM-TTS are provided in Appendix Table~\ref{tab:compound_attack}. Higher is better.}
\label{fig:compound_attack}
\end{figure*}

% 图~\ref{fig:compound_attack}展示了GAMW和四种基线方法在CosyVoice2上的八种复合扰动结果,CosyVoice2和GLM-TTS的完整数值见附录表~\ref{tab:compound_attack}，每个平均值由八种复合扰动等权计算。
Figure~\ref{fig:compound_attack} compares GAMW with four baselines under eight compound perturbations on CosyVoice2. Complete numerical results for CosyVoice2 and GLM-TTS are reported in Appendix Table~\ref{tab:compound_attack}, with each average computed equally over the eight conditions.

% 如图~\ref{fig:compound_attack}所示，GAMW在CosyVoice2的全部八种复合扰动中均取得最佳结果，平均ACC达到0.9145，较最强基线提高23.71个百分点。附录表~\ref{tab:compound_attack}进一步显示，GLM-TTS呈现一致趋势：GAMW在全部八种条件下同样排名第一，平均ACC达到0.9071，较最强基线提高23.31个百分点。即使在EnCodec、Opus与量化组成的三级级联下，GAMW仍分别保持0.8332和0.8218的ACC。上述结果表明，GAMW能够在连续多步骤处理后保持较强的载荷可恢复性。
As shown in Fig.~\ref{fig:compound_attack}, GAMW ranks first in all eight compound conditions on CosyVoice2, achieving an average ACC of 0.9145 and outperforming the strongest baseline by 23.71 percentage points. Appendix Table~\ref{tab:compound_attack} shows the same trend on GLM-TTS: GAMW again ranks first in every condition, with an average ACC of 0.9071 and a 23.31-point advantage over the strongest baseline. Even under the three-stage EnCodec, Opus, and quantization cascade, GAMW retains ACCs of 0.8332 and 0.8218 on the two backbones. These results demonstrate strong payload recoverability after sequential multi-step processing.

\subsection{Ablation Study}
% 消融实验从损失函数和嵌入位置两个维度分析GAMW的设计选择。两组实验均采用CosyVoice2和16比特水印。
Ablation experiments analyze two design choices in GAMW: the decoding loss formulation and the embedding position. Both studies use CosyVoice2 with a 16-bit watermark.

\subsubsection{Loss Function}
% 为分析Focal调制和逐比特加权（PBW）对训练初期优化的作用，我们采用$2\times2$消融设计，比较标准BCE、Focal BCE、BCE+PBW以及完整的Focal+PBW目标。四个变体采用相同的模型结构、训练数据、训练环境，仅改变损失函数。我们在第一个epoch内每隔100个训练步使用同一验证集评估一次，同时记录epoch结束时的结果。
To examine the roles of focal modulation and per-bit weighting (PBW) in early optimization, we use a $2\times2$ design comparing standard BCE, focal BCE, BCE+PBW, and the complete focal+PBW objective. The four variants use the same model architecture, training data, and training environment, differing only in the loss function. During the first epoch, we evaluate each variant on the same validation set every 100 training steps and also record the epoch-end result.

\begin{figure*}[htbp]
\centering
\includegraphics[width=\textwidth]{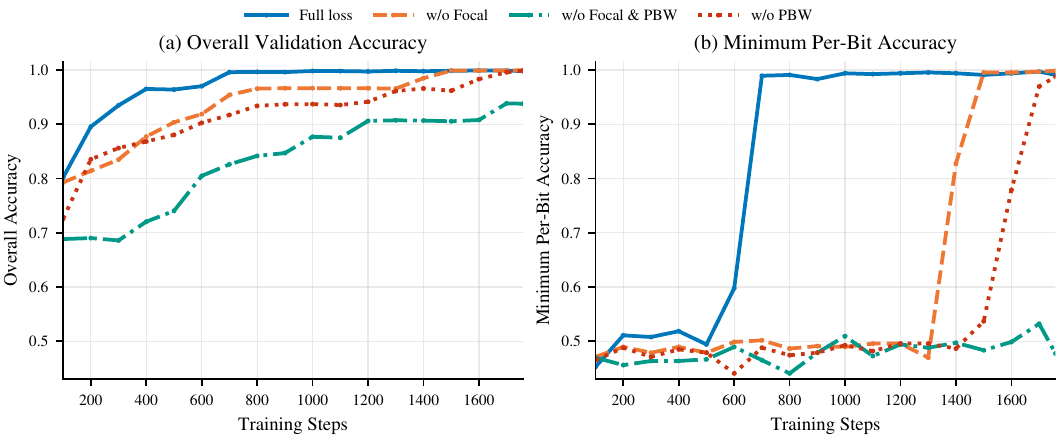}
% 第一个无扰动预热epoch内不同解码目标的验证性能变化。(a)固定验证集上的整体水印解码准确率；(b)16个水印比特中的最低逐比特准确率。训练期间每隔100步在同一验证集上评估一次，并包含该epoch结束时的验证结果。
\caption{Validation performance of different decoding objectives during the first clean warm-up epoch. (a) Overall watermark decoding accuracy on a fixed validation set. (b) Minimum per-bit accuracy across the 16 watermark bits. Evaluation is performed on the same validation set every 100 training steps, with the epoch-end result also included.}
\label{fig:loss_convergence}
\end{figure*}

% 如图~\ref{fig:loss_convergence}所示，去除Focal或PBW都会延迟整体验证准确率与最低逐比特准确率的同步提升。以整体验证ACC不低于0.99且最低逐比特ACC不低于0.95为统一判据，完整目标在第700步达到该标准，而去除Focal和去除PBW的变体分别在第1500步和第1700步达到；标准BCE在第一个epoch内始终未达到。上述结果表明，两个组件均有助于训练初期优化，联合使用时能够最快实现整体性能与困难比特恢复的均衡收敛。
As shown in Fig.~\ref{fig:loss_convergence}, removing either focal modulation or PBW delays the simultaneous improvement of overall validation accuracy and minimum per-bit accuracy. Using an overall validation ACC of at least 0.99 and a minimum per-bit ACC of at least 0.95 as a common criterion, the complete objective reaches this level at step 700, whereas the variants without focal modulation and without PBW reach it at steps 1,500 and 1,700, respectively. Standard BCE does not satisfy the criterion within the first epoch. These results show that both components contribute to early optimization, and their joint use achieves the fastest balanced convergence of overall performance and difficult-bit recovery.

\subsubsection{Embedding Position}
% 为比较mel频谱域与最终波形域嵌入，我们在CosyVoice2的16比特设置下采用与两种表示匹配的编码器和提取器，并保持训练数据与评测协议一致。波形域变体以波形L1损失替代mel域MSE。表~\ref{tab:ablation_pos}报告音质与鲁棒性，其中$\mathrm{ACC}_{\mathrm{rob}}$为15种单一扰动下ACC的等权平均值。
To compare mel- and waveform-domain embedding, we use representation-specific encoders and extractors on CosyVoice2 with a 16-bit watermark, while keeping the training data and evaluation protocol aligned. The waveform variant replaces mel-domain MSE with waveform L1 loss. Table~\ref{tab:ablation_pos} reports quality and robustness, where $\mathrm{ACC}_{\mathrm{rob}}$ averages ACC equally over 15 individual perturbations.

\begin{table}[htbp]
\centering
\small
\setlength{\tabcolsep}{2.5pt}
\caption{Ablation of embedding position on CosyVoice2 with a 16-bit watermark.}
\label{tab:ablation_pos}
\begin{tabular}{lccccc}
\hline
\textbf{Position} & \textbf{PESQ} & \textbf{ViSQOL} & \textbf{MOS} & \textbf{ACC}$\mathbf{_{\text{clean}}}$ & \textbf{ACC}$\mathbf{_{\text{rob}}}$ \\
\hline
Waveform    & 2.1911 & 3.8863 & 3.2943 & 0.9991 & 0.9048 \\
Mel (ours)  & 4.2914 & 4.9721 & 4.0790 & 0.9999 & 0.9795 \\
\hline
\end{tabular}
\end{table}

% 表~\ref{tab:ablation_pos}表明，两种变体在无扰动下均可近乎完整地恢复水印。波形域嵌入的扰动后平均ACC为0.9048，但PESQ、ViSQOL和MOS仅为2.1911、3.8863和3.2943。mel域嵌入将平均ACC提高至0.9795，并将三项音质指标保持在4.2914、4.9721和4.0790，在所评估的两种实现中取得了更好的不可感知性—鲁棒性权衡。
Table~\ref{tab:ablation_pos} shows nearly perfect clean recovery for both variants. Waveform-domain embedding attains 0.9048 post-perturbation ACC, but only 2.1911 PESQ, 3.8863 ViSQOL, and 3.2943 MOS. Mel-domain embedding raises ACC to 0.9795 while maintaining these quality metrics at 4.2914, 4.9721, and 4.0790, providing a better imperceptibility--robustness trade-off between the two evaluated implementations.

\begin{figure}[htbp]
\centering
\includegraphics[width=\columnwidth]{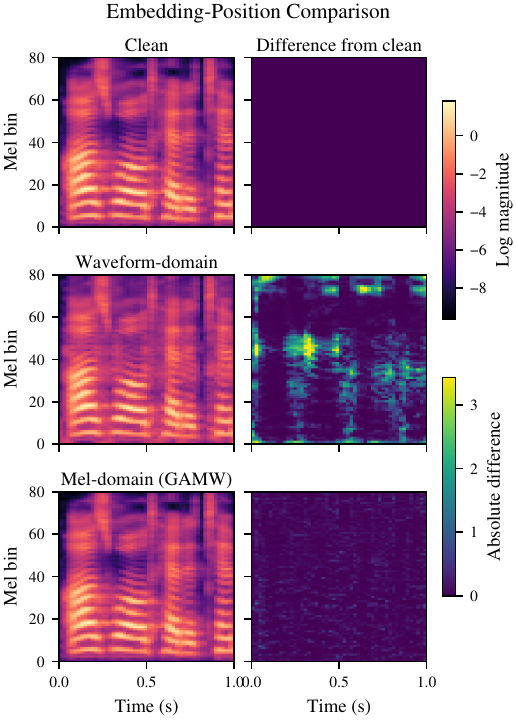}
% CosyVoice2上波形域与mel域嵌入的频谱可视化。图中比较干净语音、两种含水印语音的mel频谱及其相对于干净频谱的绝对差值。
\caption{Spectral comparison of waveform-domain and mel-domain embedding on CosyVoice2. The figure shows the mel-spectrograms of clean and watermarked speech together with the absolute differences from the clean spectrogram.}
\label{fig:embedding_position_visualization}
\end{figure}

% 我们进一步可视化干净语音及两种含水印语音的mel频谱。波形域频谱从含水印波形中重新提取；图中同时给出相对干净频谱的绝对差值，并对同类面板使用一致的颜色范围。
We further visualize mel-spectrograms of clean and watermarked speech. The waveform-domain spectrogram is re-extracted from its output audio; absolute differences from the clean spectrogram are shown with consistent color ranges across each panel type.

% 图~\ref{fig:embedding_position_visualization}显示，两种方法均保留了主要时频结构，但波形域嵌入产生更明显的频谱偏差，GAMW的修改更弱且随局部语音内容分布。这与表~\ref{tab:ablation_pos}的音质和鲁棒性结果一致。
Figure~\ref{fig:embedding_position_visualization} shows that both methods preserve the dominant time--frequency structure, but waveform-domain embedding causes larger spectral deviations, whereas GAMW makes weaker, content-dependent modifications. This agrees with the quality and robustness results in Table~\ref{tab:ablation_pos}.

\section{Conclusion}
% 本文提出GAMW，一种面向TTS生成语音溯源的无需参考多比特mel频谱水印框架。GAMW在声学模型与声码器之间嵌入内容自适应水印残差，使水印与语音内容共同经过冻结声码器生成波形，并结合Focal调制与动态逐比特加权改善训练初期的解码学习。在CosyVoice2和GLM-TTS上分别训练和评估的结果表明，GAMW能够在保持语音质量和较高载荷无扰动解码能力的同时，在单一与复合扰动下取得较高的平均解码准确率。嵌入位置消融进一步表明，相较于最终波形域嵌入，mel域嵌入在语音保真度和扰动后水印恢复之间取得了更好的经验性平衡。上述结果验证了将mel接口用于TTS生成语音主动溯源的可行性。
This paper presented GAMW, a reference-free multi-bit mel-spectrogram watermarking framework for TTS provenance. GAMW inserts content-adaptive watermark residuals between the acoustic model and vocoder, allowing the watermark and speech content to be rendered jointly by a frozen vocoder, and combines focal modulation with dynamic per-bit weighting to facilitate early decoding optimization. Results from separately trained models on CosyVoice2 and GLM-TTS show that GAMW maintains speech quality and reliable clean decoding at higher payloads while achieving high average decoding accuracy under individual and compound perturbations. The embedding-position ablation shows that mel-domain insertion provides a better empirical balance between speech fidelity and post-perturbation watermark recovery than waveform-domain insertion. These findings support the feasibility of using the mel interface for active provenance of TTS-generated speech.

\section*{Limitations}
% 本研究仍存在若干局限。GAMW仅在CosyVoice2和GLM-TTS上进行验证，两者均包含mel频谱生成与神经声码器阶段，且水印模型针对每个基座分别训练，因此当前结果不能说明其能够直接迁移至不使用mel接口的TTS架构，也不代表单个水印模型具备跨基座泛化能力。实验覆盖了多类单一扰动和预设复合扰动，但神经音频编解码仅测试EnCodec，尚未系统考察更多编解码器、未见扰动以及面向提取器的自适应移除、伪造或覆盖攻击。音质评估主要依赖客观指标和UTMOS预测，仍需通过主观听测进一步验证感知透明性。未来将面向更多TTS生成流程和声码器扩展评测，并研究更高效的基座适配与面向自适应攻击的安全训练。
This study has several limitations. GAMW is evaluated only on CosyVoice2 and GLM-TTS, both of which include mel-spectrogram generation and neural vocoding, and the watermark model is trained separately for each backbone. The results therefore do not establish direct applicability to TTS architectures without a mel interface or cross-backbone generalization of a single watermark model. Although the evaluation covers diverse individual and predefined compound perturbations, EnCodec is the only neural audio codec tested, and broader codec families, unseen perturbations, and adaptive removal, forgery, or overwriting attacks targeting the extractor remain unexplored. Audio quality is assessed mainly through objective metrics and UTMOS predictions, so subjective listening tests are still needed to confirm perceptual transparency. Future work will extend evaluation to more TTS pipelines and vocoders and investigate efficient backbone adaptation and security training against adaptive attacks.

\bibliography{custom}

\appendix

\section{Implementation Details}
\label{app:implementation_details}

\subsection{Data Preparation}
% 训练与测试清单均采用JSONL格式，每条记录包含语句标识、文本、参考语音路径和说话人信息。预处理首先移除字段缺失、音频文件不存在或超过时长限制的记录，并保证LibriSpeech官方测试集与LJSpeech人工划分测试集不参与训练。对于每个TTS骨干，我们使用其自身的文本前端、语音提示编码和声学模型生成80维mel频谱，并按语句标识缓存为张量文件。因此，CosyVoice2和GLM-TTS使用相同的文本—语音样本划分，但分别使用与各自生成过程匹配的mel表示，而不是在两个骨干之间共享同一份mel缓存。
Training and test manifests use JSONL records containing an utterance identifier, text, a reference-speech path, and speaker information. Preprocessing removes records with missing fields, unavailable audio, or excessive duration, and excludes the official LibriSpeech test set and the manually held-out LJSpeech test split from training. For each TTS backbone, we use its own text frontend, speech-prompt encoding, and acoustic model to generate 80-bin mel-spectrograms, which are cached as tensors indexed by utterance identifier. CosyVoice2 and GLM-TTS therefore use the same text--speech data split but separate mel caches matched to their respective generation processes.

% 水印训练采用短时mel片段。缓存或加载阶段保留生成mel频谱的前50帧；若同一批次中的有效长度不同，整理函数将各样本截取到该批次的最短长度后再拼接。该设置使编码器、冻结声码器和提取器接收形状一致的批量输入，并使训练聚焦于约一秒的短时水印恢复。完整语音仍用于需要整句波形的音质评估，避免将独立声码化的短mel前缀与完整音频质量直接比较。
Watermark training uses short mel segments. During caching or loading, we retain the first 50 frames of each generated mel-spectrogram; when valid lengths differ within a mini-batch, the collator truncates all samples to the shortest length before concatenation. This produces shape-consistent batched inputs for the encoder, frozen vocoder, and extractor, and focuses training on approximately one-second watermark recovery. Complete utterances are retained for audio-quality evaluation that requires full waveforms, avoiding direct comparison between independently vocoded short mel prefixes and full-audio quality.

\subsection{Training Protocol}
% 每个训练步骤为批次中的每个样本独立生成随机多比特载荷。训练时将原载荷及其逐比特取反版本与同一mel输入配对，以增加比特取值的平衡性；该扩展仅用于训练，验证和测试直接使用随机载荷。TTS文本生成器、声学模型和声码器参数保持冻结，仅优化水印编码器与提取器。GAMW针对CosyVoice2和GLM-TTS分别训练，不共享水印模型参数。
At each training step, an independent random multi-bit payload is generated for every sample in the mini-batch. During training, each payload and its bitwise complement are paired with the same mel input to improve bit-value balance; this expansion is used only for training, while validation and testing use random payloads directly. The TTS token generator, acoustic model, and vocoder remain frozen, and only the watermark encoder and extractor are optimized. GAMW is trained separately for CosyVoice2 and GLM-TTS without sharing watermark-model parameters.

% 模型使用AdamW优化器和余弦学习率调度训练50个epoch，初始学习率设为$5\times10^{-4}$，每张GPU的批次大小为32，并通过分布式数据并行在6张NVIDIA RTX 3090和2张RTX 4090 GPU上联合训练。总损失中的解码损失和mel保真度损失权重分别设为$\lambda_{\mathrm{dec}}=1.0$和$\lambda_{\mathrm{mel}}=0.2$；Focal调制参数设为$\gamma=2$，逐比特加权系数设为$\lambda_w=10$。训练开始阶段仅优化干净语音上的水印恢复与mel保真度；预热结束后，以0.7的概率对每个样本施加在线数据增强。每个批次至多从预设增强池随机选择一种扰动，再通过样本级随机掩码决定批内接受该扰动的语音。表~\ref{tab:training_augmentation}列出了训练增强参数，各扰动的具体定义见附录~\ref{app:perturbation_settings}。不可微操作采用直通估计器，使其能够参与端到端优化。不同载荷容量沿用相同的模型和数据流程，仅改变水印长度及相应训练配置。
Models are optimized for 50 epochs using AdamW with a cosine learning-rate schedule, an initial learning rate of $5\times10^{-4}$, and a per-GPU batch size of 32. Training uses distributed data parallelism across six NVIDIA RTX 3090 GPUs and two RTX 4090 GPUs. The decoding and mel-fidelity loss weights are set to $\lambda_{\mathrm{dec}}=1.0$ and $\lambda_{\mathrm{mel}}=0.2$, respectively; the focal modulation parameter is set to $\gamma=2$, and the per-bit weighting coefficient is set to $\lambda_w=10$. The initial warm-up phase optimizes watermark recovery and mel fidelity on clean speech. After warm-up, online augmentation is applied independently to each sample with a probability of 0.7. At most one perturbation is sampled from the predefined augmentation pool per mini-batch, and a sample-level random mask determines which utterances receive it. Table~\ref{tab:training_augmentation} lists the training-time parameters, and Appendix~\ref{app:perturbation_settings} defines the perturbations. A straight-through estimator enables non-differentiable operations to participate in end-to-end optimization. Different payload capacities follow the same model and data pipeline, with the watermark length and corresponding training configuration adjusted as needed.

\begin{table}[htbp]
\centering
\small
\setlength{\tabcolsep}{3pt}
% 训练阶段在线数据增强所使用的参数。集合表示每次采样的候选值，区间表示均匀采样范围。
\caption{Training-time augmentation parameters. Sets denote discrete sampling choices, and intervals denote uniform sampling ranges.}
\label{tab:training_augmentation}
\begin{tabular}{l l}
\hline
\textbf{Perturbation} & \textbf{Training parameter} \\
\hline
Gaussian noise & SNR $\in\{10,20,30\}$ dB \\
Background noise & SNR $\in\{10,20,30\}$ dB \\
Low-pass filtering & Normalized cutoff $=0.20$ \\
High-pass filtering & Normalized cutoff $=0.05$ \\
Smoothing & Window $\in\{6,10,14,18\}$ samples \\
Echo & Delay $\in[0.1,0.5]$ s; gain $\in[0.1,0.5]$ \\
EnCodec & Bandwidth $=24$ kbps \\
Time stretching & Speed $\in\{0.9,1.1\}$ \\
\hline
\end{tabular}
\end{table}

\section{Perturbation Definitions and Parameters}
\label{app:perturbation_settings}

% 本节给出鲁棒性实验中各扰动的实现定义与论文报告参数。CosyVoice2和GLM-TTS采用相同的扰动函数、参数和执行顺序。单一扰动由同一份加水印音频缓存分别施加；复合扰动则按预设管线从左至右依次执行。所有操作的输出均在水印提取前裁剪或补零至输入长度。测试实现支持更宽的辅助参数扫描，但正文与附录结果表仅报告表~\ref{tab:attack_definitions}中列出的设置。
This section specifies the implementation and reported parameters of the perturbations used in the robustness evaluation. CosyVoice2 and GLM-TTS use the same perturbation functions, parameters, and operation order. Individual perturbations are applied separately to the same cache of watermarked audio, whereas compound perturbations execute their predefined pipelines from left to right. Outputs are cropped or zero-padded to the input length before watermark extraction. The evaluation code supports a wider auxiliary parameter sweep, but the main-text and appendix result tables report only the settings listed in Table~\ref{tab:attack_definitions}.

\begin{table*}[htbp]
\centering
\small
\setlength{\tabcolsep}{4pt}
\caption{Definitions and reported parameters of the individual perturbations.}
\label{tab:attack_definitions}
\begin{tabular}{p{0.17\textwidth}p{0.52\textwidth}p{0.23\textwidth}}
\hline
\textbf{Perturbation} & \textbf{Definition} & \textbf{Reported parameter(s)} \\
\hline
Gaussian noise & Add zero-mean Gaussian noise scaled according to the signal power and target SNR. & SNR $\in\{10,20,30\}$ dB \\
Background noise & Repeat or truncate a real background-noise recording to the utterance length and scale it to the target SNR before mixing. & SNR $\in\{10,20,30\}$ dB \\
Low-pass filtering & Apply the Julius low-pass filter to the waveform. & Normalized cutoff $=0.2$ \\
High-pass filtering & Apply the Julius high-pass filter to the waveform. & Normalized cutoff $=0.1$ \\
MP3 compression & Encode and decode the waveform using MP3 compression with a fixed bitrate. & Bitrate $=16$ kbps \\
Opus compression & Encode and decode the waveform using Opus with a fixed bitrate and encoder complexity. & Bitrate $=16$ kbps; complexity $=1$ \\
EnCodec & Encode and decode the waveform using the 24-kHz EnCodec model. & Bandwidth $=12$ kbps \\
Smoothing & Apply a moving-average filter to the waveform. & Window size $=22$ samples \\
Echo & Add one delayed and attenuated copy of the waveform and normalize the result to the original amplitude range. & Delay $=0.3$ s; amplitude $=0.4$ \\
Quantization & Min--max normalize the waveform, uniformly quantize it, and map it back to its original amplitude range. & $16$ levels (4-bit) \\
Time stretching & Perform pitch-preserving time stretching and crop or zero-pad the result to the original length. & Speed factor $=1.1$ \\
\hline
\end{tabular}
\end{table*}

% 高斯噪声根据每条语音的平均信号功率生成；背景噪声采用Torchaudio提供的VOiCES样例录音，并在混合前重复或截断至目标长度。低通和高通参数是直接传递给Julius滤波器的归一化截止值。MP3和Opus的比特率分别以kbps和bps传入实现，表~\ref{tab:attack_definitions}统一换算为kbps。EnCodec使用\texttt{facebook/encodec\_24khz}模型及12~kbps目标带宽。量化采用16个均匀级别，对应4比特表示；时间拉伸使用1.1倍速率并保持最终样本数不变。
Gaussian noise is generated from the mean signal power of each utterance. Background noise uses the VOiCES example recording distributed through Torchaudio and is repeated or truncated before mixing. The low- and high-pass parameters are normalized cutoff values passed directly to the Julius filters. MP3 and Opus bitrates are passed to their implementations in kbps and bps, respectively, and are both reported in kbps in Table~\ref{tab:attack_definitions}. EnCodec uses the \texttt{facebook/encodec\_24khz} model with a target bandwidth of 12~kbps. Quantization uses 16 uniform levels, corresponding to a 4-bit representation, and time stretching uses a factor of 1.1 while preserving the final sample count.

\begin{table*}[htbp]
\centering
\small
\setlength{\tabcolsep}{4pt}
\caption{Definitions and parameters of the compound-perturbation pipelines. Operations are applied from left to right.}
\label{tab:compound_definitions}
\begin{tabular}{p{0.30\textwidth}p{0.64\textwidth}}
\hline
\textbf{Pipeline} & \textbf{Ordered operations and parameters} \\
\hline
Gaussian noise + background noise & Gaussian noise (20 dB) $\rightarrow$ background noise (20 dB) \\
EnCodec + Opus & EnCodec (12 kbps) $\rightarrow$ Opus (16 kbps, complexity 1) \\
Echo + Gaussian noise & Echo (0.3 s, amplitude 0.4) $\rightarrow$ Gaussian noise (20 dB) \\
EnCodec + MP3 & EnCodec (12 kbps) $\rightarrow$ MP3 (16 kbps) \\
Time stretching + smoothing + Opus & Time stretching (1.1) $\rightarrow$ smoothing (22 samples) $\rightarrow$ Opus (16 kbps, complexity 1) \\
Low-pass filtering + Gaussian noise + EnCodec & Low-pass filtering (0.2) $\rightarrow$ Gaussian noise (20 dB) $\rightarrow$ EnCodec (12 kbps) \\
Echo + background noise + MP3 & Echo (0.3 s, amplitude 0.4) $\rightarrow$ background noise (20 dB) $\rightarrow$ MP3 (16 kbps) \\
EnCodec + Opus + quantization & EnCodec (12 kbps) $\rightarrow$ Opus (16 kbps, complexity 1) $\rightarrow$ quantization (16 levels) \\
\hline
\end{tabular}
\end{table*}

% 复合扰动不重新采样参数，而是固定使用表~\ref{tab:compound_definitions}中的值，以确保所有方法和两个TTS骨干经历完全相同的处理。每一步均接收前一步的输出，因此表中顺序属于实验定义的一部分。该设置覆盖两级和三级组合，并包含噪声叠加、传统压缩、神经音频编解码、滤波、回声、平滑、时间拉伸和量化等不同类型的连续处理。
Compound perturbations do not resample parameters; they use the fixed values in Table~\ref{tab:compound_definitions} so that every method and both TTS backbones undergo identical processing. Each operation receives the output of the preceding operation, making the listed order part of the experimental definition. The pipelines cover two- and three-stage combinations involving additive noise, conventional compression, neural audio codec processing, filtering, echo, smoothing, time stretching, and quantization.

\section{Complete Individual-Perturbation Results}
\label{app:single_results}

% 本节给出正文表~\ref{tab:single_attack}所对应的16比特完整单一扰动结果。评测覆盖加性高斯噪声、背景噪声、滤波、传统有损压缩、EnCodec神经音频编解码重合成以及常见信号处理。对于高斯噪声和背景噪声，分别报告10、20和30~dB信噪比；其余扰动各采用正文实验设置中的固定参数。所有方法均使用相同的测试音频、扰动参数和指标实现进行评估。表~\ref{tab:single_attack_full}同时报告CosyVoice2和GLM-TTS上的ACC，每个骨干的平均值由全部15种扰动条件等权计算。
This section provides the complete individual-perturbation results for the 16-bit setting corresponding to Table~\ref{tab:single_attack} in the main text. The evaluation covers additive Gaussian noise, background noise, filtering, conventional lossy compression, neural audio codec re-synthesis with EnCodec, and common signal processing. Gaussian and background noise are each evaluated at SNRs of 10, 20, and 30~dB, while each remaining perturbation uses the fixed parameter specified by the main experimental protocol. All methods are evaluated using the same test audio, perturbation parameters, and metric implementation. Table~\ref{tab:single_attack_full} reports ACC on both CosyVoice2 and GLM-TTS, with each backbone average computed equally over all 15 conditions.

\begin{table*}[htbp]
\centering
\small
\setlength{\tabcolsep}{3pt}
\caption{Complete decoding accuracy under individual perturbations with a 16-bit payload on CosyVoice2 and GLM-TTS. The Average row is computed over all 15 conditions for each backbone. Higher is better.}
\label{tab:single_attack_full}
\begin{tabular}{lccccc}
\hline
\textbf{Perturbation} & \textbf{WavMark} & \textbf{AudioSeal} & \textbf{SilentCipher} & \textbf{TimbreWM} & \textbf{GAMW} \\
\hline
\multicolumn{6}{c}{\textbf{CosyVoice2}} \\
\hline
Gaussian noise (SNR = 10 dB)       & 0.5008 & 0.6711 & 0.1669 & 0.6777 & 0.9837 \\
Gaussian noise (SNR = 20 dB)       & 0.5946 & 0.8911 & 0.1751 & 0.9054 & 0.9977 \\
Gaussian noise (SNR = 30 dB)       & 0.8572 & 0.9876 & 0.2370 & 0.9922 & 0.9996 \\
Background noise (SNR = 10 dB)     & 0.6176 & 0.7256 & 0.2343 & 0.7224 & 0.9663 \\
Background noise (SNR = 20 dB)     & 0.8580 & 0.8991 & 0.2635 & 0.8918 & 0.9963 \\
Background noise (SNR = 30 dB)     & 0.9817 & 0.9842 & 0.3327 & 0.9900 & 0.9995 \\
Low-pass filtering                 & 0.9875 & 1.0000 & 0.8191 & 0.9976 & 0.9997 \\
High-pass filtering                & 0.9999 & 0.3253 & 0.3108 & 0.9672 & 0.9575 \\
MP3 compression                    & 0.7123 & 0.9850 & 0.2273 & 0.9858 & 0.9716 \\
Opus compression                   & 0.7809 & 0.8323 & 0.2115 & 0.9822 & 0.9709 \\
EnCodec                            & 0.4454 & 0.3691 & 0.1698 & 0.6045 & 0.9670 \\
Smoothing                          & 0.8953 & 0.8750 & 0.2701 & 0.9081 & 0.9951 \\
Echo                               & 0.8332 & 0.9608 & 0.2012 & 0.9988 & 0.9967 \\
Quantization                       & 0.4673 & 0.7189 & 0.1835 & 0.7637 & 0.9816 \\
Time stretching                    & 0.9600 & 0.9980 & 0.3081 & 0.9998 & 0.9981 \\
\hline
Average                            & 0.7661 & 0.8149 & 0.2740 & 0.8925 & 0.9854 \\
\hline
\multicolumn{6}{c}{\textbf{GLM-TTS}} \\
\hline
Gaussian noise (SNR = 10 dB)       & 0.5091 & 0.6198 & 0.1771 & 0.6839 & 0.9744 \\
Gaussian noise (SNR = 20 dB)       & 0.6053 & 0.8442 & 0.1753 & 0.9010 & 0.9944 \\
Gaussian noise (SNR = 30 dB)       & 0.8425 & 0.9699 & 0.2301 & 0.9862 & 0.9984 \\
Background noise (SNR = 10 dB)     & 0.5964 & 0.6715 & 0.2334 & 0.7029 & 0.9603 \\
Background noise (SNR = 20 dB)     & 0.8054 & 0.8421 & 0.2479 & 0.8596 & 0.9916 \\
Background noise (SNR = 30 dB)     & 0.9542 & 0.9670 & 0.2878 & 0.9763 & 0.9983 \\
Low-pass filtering                 & 0.9904 & 1.0000 & 0.7871 & 0.9987 & 0.9998 \\
High-pass filtering                & 1.0000 & 0.3318 & 0.3152 & 0.9731 & 0.9884 \\
MP3 compression                    & 0.8038 & 0.9957 & 0.2533 & 0.9934 & 0.9814 \\
Opus compression                   & 0.8227 & 0.8432 & 0.2272 & 0.9907 & 0.9622 \\
EnCodec                            & 0.4399 & 0.3712 & 0.1584 & 0.5971 & 0.9420 \\
Smoothing                          & 0.8619 & 0.8535 & 0.2892 & 0.9244 & 0.9799 \\
Echo                               & 0.7485 & 0.9561 & 0.2208 & 0.9998 & 0.9970 \\
Quantization                       & 0.4683 & 0.6623 & 0.2185 & 0.7634 & 0.9630 \\
Time stretching                    & 0.9662 & 0.9998 & 0.3186 & 1.0000 & 0.9989 \\
\hline
Average                            & 0.7610 & 0.7952 & 0.2760 & 0.8900 & 0.9820 \\
\hline
\end{tabular}
\end{table*}

\section{Complete Compound-Perturbation Results}
\label{app:compound_results}

% 本节给出正文图~\ref{fig:compound_attack}所对应的16比特完整复合扰动结果。复合扰动由两种或三种处理按表中顺序依次施加，用于模拟语音在传播、压缩和编辑过程中经历的连续处理。所有方法均在相同的音频、扰动类型、参数和处理顺序下评估，结果以水印解码准确率（ACC）报告。表~\ref{tab:compound_attack}同时列出CosyVoice2和GLM-TTS上的结果，平均值由每个骨干对应的八种复合扰动等权计算。
This section provides the complete compound-perturbation results for the 16-bit setting corresponding to Fig.~\ref{fig:compound_attack} in the main text. Each pipeline applies two or three operations sequentially in the order listed in the table, simulating successive processing during speech transmission, compression, and editing. All methods are evaluated using the same audio, perturbation types, parameters, and operation order, and performance is reported as watermark decoding accuracy (ACC). Table~\ref{tab:compound_attack} reports results on both CosyVoice2 and GLM-TTS, with the average for each backbone computed equally over its eight compound perturbations.

\begin{table*}[htbp]
\centering
\small
\setlength{\tabcolsep}{3pt}
\caption{Complete decoding accuracy under compound perturbations with a 16-bit payload. ACC (higher is better).}
\label{tab:compound_attack}
\begin{tabular}{lccccc}
\hline
\textbf{Perturbation pipeline} & \textbf{WavMark} & \textbf{AudioSeal} & \textbf{SilentCipher} & \textbf{TimbreWM} & \textbf{GAMW} \\
\hline
\multicolumn{6}{c}{\textbf{CosyVoice2}} \\
\hline
Gaussian noise + background noise             & 0.5925 & 0.8136 & 0.1748 & 0.8192 & 0.9949 \\
EnCodec + Opus                                & 0.4694 & 0.3681 & 0.1756 & 0.5910 & 0.8981 \\
Echo + Gaussian noise                         & 0.5610 & 0.8170 & 0.1824 & 0.8693 & 0.9890 \\
EnCodec + MP3                                 & 0.4530 & 0.3666 & 0.1718 & 0.5656 & 0.8490 \\
Time stretching + smoothing + Opus            & 0.4968 & 0.5647 & 0.1787 & 0.7751 & 0.8865 \\
Low-pass filtering + Gaussian noise + EnCodec & 0.5029 & 0.3663 & 0.1801 & 0.5648 & 0.9575 \\
Echo + background noise + MP3                 & 0.4810 & 0.5296 & 0.1660 & 0.7084 & 0.9078 \\
EnCodec + Opus + quantization                 & 0.4704 & 0.4598 & 0.1891 & 0.5255 & 0.8332 \\
\hline
Average                                        & 0.5034 & 0.5357 & 0.1773 & 0.6774 & 0.9145 \\
\hline
\multicolumn{6}{c}{\textbf{GLM-TTS}} \\
\hline
Gaussian noise + background noise             & 0.5991 & 0.7428 & 0.1793 & 0.8064 & 0.9947 \\
EnCodec + Opus                                & 0.4685 & 0.3590 & 0.1705 & 0.5823 & 0.8848 \\
Echo + Gaussian noise                         & 0.5525 & 0.7431 & 0.1938 & 0.8722 & 0.9858 \\
EnCodec + MP3                                 & 0.4628 & 0.3599 & 0.1578 & 0.5607 & 0.8397 \\
Time stretching + smoothing + Opus            & 0.4930 & 0.4958 & 0.1931 & 0.7774 & 0.8869 \\
Low-pass filtering + Gaussian noise + EnCodec & 0.5077 & 0.3641 & 0.1842 & 0.5551 & 0.9511 \\
Echo + background noise + MP3                 & 0.4774 & 0.4988 & 0.1628 & 0.7220 & 0.8919 \\
EnCodec + Opus + quantization                 & 0.4661 & 0.4524 & 0.2049 & 0.5155 & 0.8218 \\
\hline
Average                                        & 0.5034 & 0.5020 & 0.1808 & 0.6740 & 0.9071 \\
\hline
\end{tabular}
\end{table*}

\section{Additional Payload--Robustness Analysis}
\label{app:payload_robustness}

% 为考察载荷增加对水印鲁棒性的影响，我们进一步在32比特下评估GAMW，并与16比特主设置进行比较。两种容量使用相同的测试样本、扰动参数和15种单一扰动条件。该补充实验仅比较GAMW在不同容量下的表现；复合扰动仍仅在16比特主设置下评估。
To examine how a larger payload affects watermark robustness, we additionally evaluate GAMW with a 32-bit payload and compare it with the primary 16-bit setting. Both payloads use the same test samples, perturbation parameters, and 15 individual-perturbation conditions. This supplementary experiment compares GAMW across payload sizes; compound perturbations are evaluated only under the primary 16-bit setting.

\begin{table*}[htbp]
\centering
\small
\setlength{\tabcolsep}{9pt}
\caption{Individual-perturbation ACC of GAMW with 16-bit and 32-bit payloads. Higher is better.}
\label{tab:payload_robustness}
\begin{tabular}{lcccc}
\hline
\multirow{2}{*}{\textbf{Perturbation}} & \multicolumn{2}{c}{\textbf{CosyVoice2}} & \multicolumn{2}{c}{\textbf{GLM-TTS}} \\
 & \textbf{16} & \textbf{32} & \textbf{16} & \textbf{32} \\
\hline
Gaussian noise (10 dB)   & 0.9837 & 0.9467 & 0.9744 & 0.9401 \\
Gaussian noise (20 dB)   & 0.9977 & 0.9896 & 0.9944 & 0.9844 \\
Gaussian noise (30 dB)   & 0.9996 & 0.9974 & 0.9984 & 0.9954 \\
Background noise (10 dB) & 0.9663 & 0.8655 & 0.9603 & 0.8713 \\
Background noise (20 dB) & 0.9963 & 0.9642 & 0.9916 & 0.9594 \\
Background noise (30 dB) & 0.9995 & 0.9921 & 0.9983 & 0.9897 \\
Low-pass filtering       & 0.9997 & 0.9986 & 0.9998 & 0.9991 \\
High-pass filtering      & 0.9575 & 0.9240 & 0.9884 & 0.9464 \\
MP3 compression          & 0.9716 & 0.9591 & 0.9814 & 0.9543 \\
Opus compression         & 0.9709 & 0.9569 & 0.9622 & 0.9450 \\
EnCodec                  & 0.9670 & 0.9157 & 0.9420 & 0.9092 \\
Smoothing                & 0.9951 & 0.9909 & 0.9799 & 0.9598 \\
Echo                     & 0.9967 & 0.9885 & 0.9970 & 0.9935 \\
Quantization             & 0.9816 & 0.9412 & 0.9630 & 0.9285 \\
Time stretching          & 0.9981 & 0.9953 & 0.9989 & 0.9984 \\
\hline
Average                  & 0.9854 & 0.9617 & 0.9820 & 0.9583 \\
\hline
\end{tabular}
\end{table*}

% 如表~\ref{tab:payload_robustness}所示，载荷从16比特增加到32比特后，CosyVoice2和GLM-TTS上的平均ACC均下降约2.37个百分点，但32比特设置下仍分别达到0.9617和0.9583。下降主要出现在低信噪比背景噪声、高通滤波和EnCodec重合成等较困难条件，而低通滤波、较高信噪比噪声和时间拉伸下的ACC仍接近1。结果表明，提高载荷会带来预期的鲁棒性损失，但GAMW在所评估的单一扰动下仍保持约0.96的平均比特准确率。
As shown in Table~\ref{tab:payload_robustness}, increasing the payload from 16 to 32 bits reduces the average ACC by approximately 2.37 percentage points on both CosyVoice2 and GLM-TTS, while the 32-bit setting still reaches 0.9617 and 0.9583, respectively. The degradation is concentrated under more challenging conditions such as low-SNR background noise, high-pass filtering, and EnCodec re-synthesis, whereas ACC remains close to 1 under low-pass filtering, higher-SNR noise, and time stretching. These results show the expected robustness cost of a larger payload while retaining approximately 0.96 average bit accuracy under the evaluated individual perturbations.

\section{Failure-Case Analysis}
\label{app:failure_cases}

\begin{figure}[htbp]
\centering
\includegraphics[width=\columnwidth]{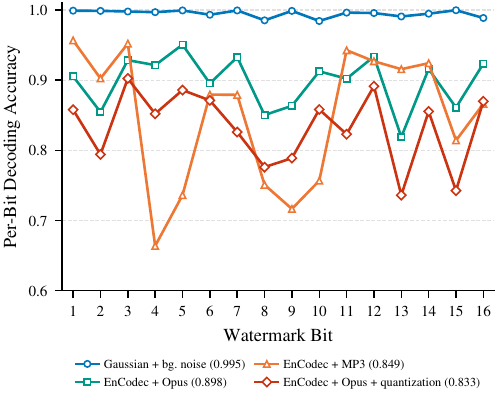}
% CosyVoice2上代表性复合扰动下的逐比特水印解码准确率。图例括号内为对应扰动下16个水印位的平均ACC。
\caption{Per-bit watermark decoding accuracy under representative compound perturbations on CosyVoice2. Values in parentheses are mean ACCs over the 16 watermark bits.}
\label{fig:failure_case_per_bit}
\end{figure}

% 为分析平均准确率无法体现的错误分布，我们选取一种相对容易的复合扰动和三种包含神经音频编解码与有损压缩的困难条件，比较GAMW在CosyVoice2上的逐比特解码准确率。图~\ref{fig:failure_case_per_bit}中的括号数值表示对应条件下16个水印位的平均ACC。
To examine error patterns obscured by average accuracy, we compare the per-bit decoding accuracy of GAMW on CosyVoice2 under one relatively easy compound perturbation and three challenging pipelines involving neural audio coding and lossy compression. The values in parentheses in Fig.~\ref{fig:failure_case_per_bit} denote the mean ACC over the 16 watermark bits for each condition.

% 在高斯噪声与背景噪声组合下，所有水印位的ACC均接近1，说明该条件未造成明显的逐比特恢复困难。相比之下，包含EnCodec和低码率压缩的复合扰动使大多数比特的准确率下降，同时部分位置出现更明显的退化。例如，EnCodec与MP3组合下第4位的ACC降至约0.66，而EnCodec、Opus与量化的三级级联对第13位和第15位影响较大。由于不同处理管线下最脆弱的比特位置并不完全一致，这些结果表明，连续有损处理既会整体削弱水印解码证据，也会形成与处理管线相关的非均匀逐比特错误。该分析用于揭示困难复合扰动下的失败模式，而不用于单独验证损失函数组件或嵌入位置的作用。
Under combined Gaussian and background noise, all watermark bits retain ACCs close to 1, indicating little bit-specific degradation. In contrast, compound perturbations involving EnCodec and low-bitrate compression reduce the accuracy of most bits, with substantially larger degradation at particular positions. For example, the ACC of Bit~4 falls to approximately 0.66 under EnCodec followed by MP3, while the three-stage EnCodec, Opus, and quantization pipeline particularly affects Bits~13 and~15. Because the most vulnerable positions differ across processing pipelines, successive lossy operations appear to both weaken watermark evidence globally and induce pipeline-dependent, non-uniform bit errors. This analysis characterizes failure patterns under difficult compound perturbations rather than independently validating the loss components or embedding position.

\section{Use of AI Assistants}
% 本研究使用AI助手提供写作辅助，包括语句润色、语法纠错和表达优化。AI助手不参与核心研究过程，包括方法设计、实验执行和结果核验。所有研究内容与论文中的最终表述均由作者审核并负责。
AI assistants were used for writing support, including sentence polishing, grammar correction, and improvements to clarity and expression. They were not involved in core research activities, including method design, experiment execution, or result verification. The authors reviewed and take responsibility for all research content and final statements in the paper.

\end{document}